\documentclass[preprint,showpacs,showkeys,aps,prb]{revtex4}
\usepackage[dvips]{graphicx}
\begin{document}

\title{
Time-Reversible Ergodic Maps and the 2015 Ian Snook Prize
}

\author{
William Graham Hoover and Carol Griswold Hoover \\
Ruby Valley Research Institute                  \\
Highway Contract 60, Box 601                    \\
Ruby Valley, Nevada 89833                       \\
}

\date{\today}


\vspace{0.1cm}

\begin{abstract}
The {\it time reversibility} characteristic of Hamiltonian mechanics has long been extended
to nonHamiltonian dynamical systems modeling nonequilibrium steady states with feedback-based
thermostats and ergostats.  Typical solutions are multifractal attractor-repellor phase-space
pairs with reversed momenta and unchanged coordinates, $(q,p)\longleftrightarrow (q,-p)$ .
Weak control of the temperature, $\propto p^2$  and its fluctuation, resulting in ergodicity,
has recently been achieved in a three-dimensional time-reversible model of a heat-conducting
harmonic oscillator.  Two-dimensional cross sections of such nonequilibrium flows can be
generated with time-reversible dissipative maps yielding \ae sthetically interesting
attractor-repellor pairs.  We challenge the reader to find and explore such time-reversible
dissipative  maps.  This challenge is the 2015 Snook-Prize Problem.

\end{abstract}

\maketitle

\section{Time-Reversible Nonequilibrium Flows}
The microscopic gist of the macroscopic Second Law is often pictured as a system's seeking out
{\it more} phase-space states.  So it seems a bit odd that the idealized nonequilibrium steady
states generated by molecular dynamics behave in the opposite way\cite{b1}.  The $(q,p)$ states
from such atomistic simulations soon become constrained to an ever-shrinking strange attractor.
The compensating additional heat-reservoir states are never seen explicitly.  Instead they are
{\it modelled} by time-reversible friction coefficients.  These coefficients control temperature
or energy and induce steady-state behavior in the system under study.

Unlike {\it real life}, the underlying Laws of physics are mostly time-reversible, which means
it is puzzling that physics gives a good accounting of real-world observations.  Time-reversible 
mechanical models ( molecular dynamics ) provide clear examples of this conundrum.  To
illustrate the irreversible behavior of such time-reversible models let us consider the simplest
case, a one-dimensional harmonic oscillator (with unit mass and force constant ) exposed to a
temperature gradient\cite{b2,b3}, $T = T(q)$ . The oscillator's motion is subject to a
time-dependent friction coefficient $\zeta(p)$ imposing weak control over the oscillator's
kinetic energy $ \langle \ K \ \rangle $ and its fluctuation,
$\langle \ K^2 \ \rangle - \langle \ K \ \rangle^2 \ $ .  This oscillator model\cite{b2}
generates a three-dimensional ``flow'' satisfying the three ordinary differential equations of
motion :
$$
\dot q = p \ ; \ \dot p = -q - \zeta [ \ Ap + B(p^3/T) \ ] \ ; \ \dot \zeta =
A[ \ (p^2/T) - 1 \ ] + B[ \ (p^4/T^2) - 3(p^2/T) \ ] \ .
$$
The superior dots in the motion equations indicate comoving time derivatives.

{\it At thermal equilibrium} where $T$ is constant the parameters $(A,B)$ are chosen to promote
the control of the second and fourth velocity moments\cite{b3} :
$$
(A,B) = ( 0.05,0.32 ) \longrightarrow \langle \ (p^2/T),(p^4/T^2) \ \rangle = (1,3) \ .
$$
From the visual standpoint this parameter choice appears to provide an ergodic coverage of
the oscillator phase space.
Liouville's continuity equation in phase space can be used to show that these motion equations
are consistent with Gibbs' canonical distribution ;
$$
f(q,p,\zeta)_{\epsilon = 0} \propto e^{-q^2/2T}e^{-p^2/2T}e^{-\zeta^2/2} \ .
$$

{\it Away from equilibrium} the heat-reservoir temperature field $T = 1 + \epsilon \tanh(q)$
is an explicit function of the oscillator coordinate $q$ and $\epsilon$ is the maximum value
of the temperature gradient.

\begin{figure}
\includegraphics[width=5.5in,angle=90.]{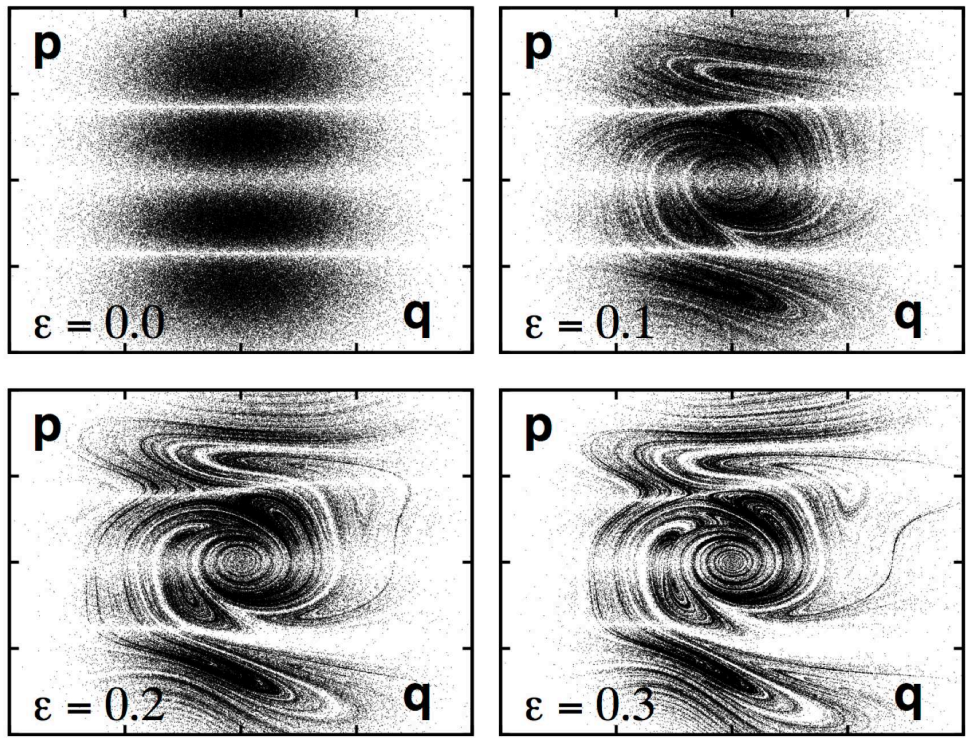}
\caption{
These cross sections can be viewed as generated by time-reversible ``dynamical maps'',
advancing the dynamics from one penetration of the $\zeta=0$ plane to the next.  These
$(q,p)$ phase-space cross sections for the conducting oscillator correspond to four values
of the maximum temperature gradient : $\epsilon = 0.0, \ 0.1, \ 0.2, \ {\rm and} \  0.3$ .
Both $q$ and $p$ range from $-4$ to $+4$ in all the cross sections.
}
\end{figure}

See {\bf Figure 1} for sample numerical $(q,p)$ cross sections of the three-dimensional flow
in $(q,p,\zeta)$ space. The $(q,p)$ points are plotted whenever $\zeta$ changes sign.  Notice
that the equations of motion for this heat-conduction model are {\it time-reversible}. To see
this start out with a solution of the equations forward in time, $q(t),p(t),\zeta(t)$ .  Then
change the signs of the momentum $p$, $\zeta$, as well as the direction of time,
$( \ +d/dt \rightarrow -d/dt \ )$ . These steps provide a new ``reversed'' solution of exactly
the same motion equations.

The nonequilibrium $(q,p)$ cross-sections with $\epsilon > 0$ shown in {\bf Figure 1} look
like inhomogenous ( or `` multifractal'' ) strange attractors.  And they are.
In these flows, as in a wide variety of time-reversible nonequilibrium steady states,
the fractal strange attractors satisfy the Second Law of Thermodynamics, with an
overall hot-to-cold heat current.  The reversed flows, topologically similar fractals,
are mechanically unstable, with exponentially growing phase volume, and with a
heat-flow direction violating the second law.  These repellor states are unobservable
numerically due to this ( Lyapunov ) instability, and can only be collected by storing,
and then reversing, a time series of attractor states\cite{b1}.

\section{Equivalent Time-Reversible Maps\cite{b4,b5}}

The construction of the flow cross-sections is equivalent to applying a time-reversible
``map'' ${\cal M}(q,p)$ from one penetration of the zero $\zeta$ plane to the next,
 $(q_{n+1},p_{n+1}) = {\cal M}(q_n,p_n) \ $.  Such two-dimensional maps provide relatively
simple ( two-dimensional rather than three ) pictures of the chaos generated by the
conducting oscillator.

Because the penetrations of the differential equations are proportional to the ``flux''
through the sampling plane $\zeta = 0$ at $(q,p,0)$ , rather than just the ``density'',
there are unoccupied white ``nullclines'' in the cross sections wherever the flux
$\dot \zeta(q,p)f(q,p,0)$ vanishes.

``Maps'' have long been used in dynamical systems theory to illustrate and explain
ergodicity ( closely approaching all of a system's states ) and Lyapunov instability
( the exponentially-fast growth of small perturbations ).  Arnold's Cat Map\cite{b6}
and the Baker Map are the best-known examples.  For clarity let us recall the action
of the Baker Map.

\begin{figure}
\includegraphics[width=5.0in,angle=90.]{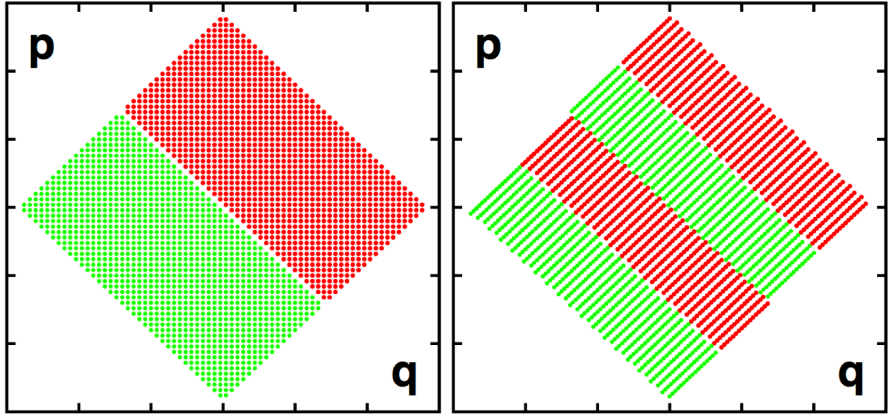}
\caption{
One iteration of the Baker Map is illustrated here for a grid of 1442 $(q,p)$
points.  The square shown at the left is first squeezed twofold in the $q=p$
direction and stretched twofold in the perpendicular direction.  After a cut
along the $q=p$ line the two halves are joined together in the new arrangement
shown at the right.  The changed spacings of the points reflect the squeezing
and stretching induced by the map.
}
\end{figure}

\section{The Rotated Time-Reversible Baker Map}
In the dynamical systems literature the discontinuous ``Baker Map'', composed
of area-preserving cutting and kneading operations, provides a simple illustration
of ``chaos'', the exponentially diverging growth induced by the mapping of
small perturbations.  The familiar ``Cat Map'' is another well-known
discontinuous mapping of the unit square into itself.  Both these map types
illustrate Lyapunov instability, $\delta_{n+1} \propto \delta_n$ , where the
proportionality constant exceeds unity, leading to divergence.

Many-to-one mappings, such as $x_{n+1} = \alpha x_n \ ; \ \alpha < 1$ , in the presence
of chaos are enough to produce the strange attractors associated with nonequilibrium
steady states.  Even one-to-one volume-preserving mappings, as in the simplest Baker
Map above, are enough to generate Lyapunov unstable chaos.  In the equilibrium
Baker Map each half of a $2\times 2$ square is mapped from a $1 \times 2$ rectangle
to two $(1/2) \times 2$ disjoint rectangles.  In order to follow our criterion for time
reversibility -- detailed in the next Section -- we use a {\it rotated} version of
the Baker Map, illustrated in {\bf Figure 2}. The mapping is as follows :                   \\

\pagebreak

\noindent
{\tt
      if(q.lt.p) then                                         \\
        qnew = +1.25d00*q - 0.75d00*p + dsqrt(1.125d00)       \\
        pnew = -0.75d00*q + 1.25d00*p - dsqrt(0.125d00)       \\
      endif                                                   \\
      if(q.gt.p) then                                         \\
        qnew = +1.25d00*q - 0.75d00*p - dsqrt(1.125d00)       \\
        pnew = -0.75d00*q + 1.25d00*p + dsqrt(0.125d00)       \\
      endif                                                   \\
}
The result of 100,000 iterations of this map is a near uniform covering of the
$2 \times 2$ square, as is shown in {\bf Figure 3}.

\begin{figure}
\includegraphics[width=5.0in,angle=90.]{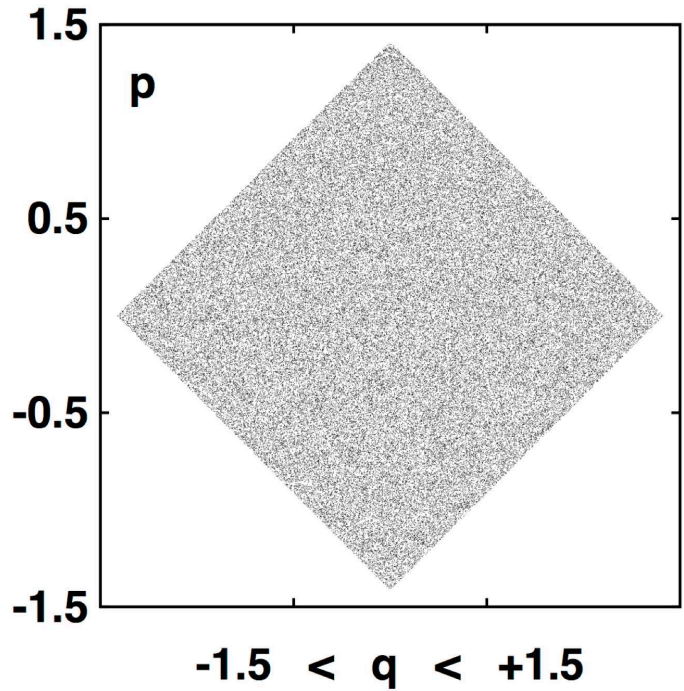}
\caption{
The uniform distribution of 100,000 iterates of the Baker Map with an initial point
$(q,p) = (0.5,0.0)$ are plotted here.  The uniform area conservation of the Map is
analogous to the equilibrium phase-volume conservation described by Liouville's
Theorem.
}
\end{figure}

The {\it dissipative} Baker Map\cite{b4,b5}, with changes in {\it area} as well as
shape, is a better model for {\it nonequilibrium} molecular dynamics and statistical
mechanics.  In those disciplines the phase-space density responds to heat transfer.
In the oscillator example above the action of a heat reservoir with temperature
$T(q)$ is represented by the time-reversible frictional force,
$- \zeta [ \ Ap + B(p^3/T) \ ]$ .  The resulting change in comoving phase
volume $\otimes$ follows from the continuity equation :
$$
(\dot \otimes/\otimes) = (\partial \dot q/\partial q) +
(\partial \dot p/\partial p) +(\partial \dot \zeta/\partial \zeta ) =
-\zeta[ \ A + 3B(p^2/T) \ ] \ .
$$
The dissipative Baker Map is deterministic, time-reversible, and likewise produces
mirror-image attractor-repellor multifractal pairs.  These are the same qualities
associated with nonequilibrium molecular dynamics algorithms ever since the early
1970s, though they passed unrecognized until the 1980s.\cite{b1}

\section{New Maps -- Ian Snook Prize for 2015}

The conducting oscillator is just one example of the deterministic, time-reversible
flows that represent a nonequilibrium steady state with a chaotic multifractal
attractor.  The time-reversed state, with the momentum, friction coefficient, and
the time all changed in sign, is an exactly similar mirror-image multifractal
structure, an unstable zero-measure repellor.  More complicated maps with these
same properties can be constructed by concatenating time-symmetric combinations of
reversible mappings $\{ \ {\cal M} \ \}$ like
$$
{\cal M}_1{\cal M}_2{\cal M}_3{\cal M}_4{\cal M}_3{\cal M}_2{\cal M}_1 \ .  
$$
where each of the mappings satisfies the time-reversibility criterion :
$$
(q,p) = {\cal TM TM}(q,p) \ ,
$$
with the four operations performed from right to left. Here ${\cal T}$ indicates
the time-reversal operation $(q,p) \longrightarrow (q,-p)$ . This reversibility
criterion states that the sequence of four steps --  [ 1 ] iterate forward;
[ 2 ] reverse velocities; [ 3 ] iterate backward; [ 4 ] reverse velocities --
returns to the original $(q,p)$ state.  The simplest time-symmetric mappings
satisfying this criterion are shears and reflections.\cite{b4} It is interesting
to note that {\it numerical} implementations almost never return {\it exactly}
to their initial state after the four-step sequence above.  In fact the {\it lack}
of an exact return is a ( somewhat misleading )\cite{b7} measure of algorithmic
accuracy\cite{b8}.

The similarities between small-system dynamics and macroscopic dissipative
behavior motivate the study of relatively simple flows and maps capable of
generating the complexity associated with irreversible strange attractors.
At the same time this complexity often exhibits a compelling beauty.
The prototypical Cat and Baker maps are relatively simple, but their discontinuities
are not at all representative of the attractor types shown in {\bf Figure 1} .
The time is ripe for a fresh look at such problems.

The Snook Prize problem for 2015 is to formulate and analyze an interesting
time-reversible ergodic map relevant to statistical mechanics, in two or three
dimensions. It is desirable that the map generates a multifractal.  Entries can
be submitted to the authors, to the Los Alamos ar$\chi$iv, or to Computational
Methods in Science and Technology. The author(s) of the most interesting entry
received prior to 1 January 2016 will receive the Prize's cash award of \$500 US .
The publisher of Computational Methods in Science and Technology ( cmst.eu ) has
generously offered an Additional Prize award of \$555.5$\dot 5$ US subject to the
Terms and Conditions published in CMST .
\pagebreak


\begin{thebibliography}{8}

\bibitem{b1}  W. G. Hoover and C. G. Hoover, {\it Simulation and Control of
              Chaotic Nonequilibrium Systems} (World Scientific Publishers,
              Singapore, 2015) .
\bibitem{b2}  H. A. Posch and Wm. G. Hoover, ``Time-Reversible Dissipative
              Attractors in Three and Four Phase-Space Dimensions'', Physical
              Review E {\bf 55}, 6803-6810 (1997).
\bibitem{b3}  W. G. Hoover, J. C. Sprott, and C. G. Hoover, ``Nonequilibrium
              Molecular Dynamics and Dynamical Systems theory for Small Systems
              with Time-Reversible Motion Equations'', Molecular Simulations
              ( in preparation, 2015 ) .
\bibitem{b4}  W. G. Hoover, O. Kum, and H. A. Posch, ``Time-Reversible Dissipative
              Ergodic Maps'', Physical Review E {\bf 53}, 2123-2129 (1996) .
\bibitem{b5}  J. Kumi\u{c}\'ak, Irreversibility in a Simple Reversible Model'',
              Physical Review E {\bf 71}, 016115 (2005) = ar$\chi$iv nlin/0510016 .
\bibitem{b6}  L. Ermann and D. L. Shepelyansky, ``Arnold Cat Map, Ulam Method,
              and Time Reversal'', Physica D {\bf 241}, 514-518 (2012) = ar$\chi$iv
              1107.0437 .
\bibitem{b7}  W. G. Hoover and C. G. Hoover, ``Comparison of Very Smooth Cell-Model
              Trajectories Using Five Symplectic and Two Runge-Kutta Integrators'',
              Computational Methods in Science and Technology {\bf 21}(to appear,
              2015) = ar$\chi$iv 1504.00620 .
\bibitem{b8}  D. Faranda, M. F. Mestre, and G. Turchetti, ``Analysis of Roundoff
              Errors with Reversibility Test as a Dynamical Indicator'', International
              Journal of Bifurcation and Chaos {\bf 22}, 1250215 (2012) = ar$\chi$iv
              1205.3060 .

\end{thebibliography}
\end{document}